\documentclass{aa}
\usepackage{graphicx}
\begin{document}
\newcommand{\etal}{{ et al. }} 
   \title{Supermassive Black Hole Masses of AGNs with Elliptical Hosts
 }

   \author{Xue-Bing Wu          
\and F.K. Liu    \and T.Z. Zhang     }

   \offprints{X.-B. Wu}

   \institute{Department of Astronomy, School of Physics, Peking University and CAS-PKU 
joint Beijing Astrophysics Center, 
Beijing 100871, China\\
              \email{wuxb@bac.pku.edu.cn; fkliu@bac.pku.edu.cn; bzty@bac.pku.edu.cn}
             }

   \date{Received 04/02/2002; accepted 11/03/2002}

   \abstract{The recently discovered tight correlation between supermassive black hole mass  and 
central velocity dispersion for both inactive and 
active galaxies suggests a possibility to estimate the
black hole mass from the measured central velocity dispersion. However, for 
 most AGNs it is difficult to measure the central velocity dispersions of their
host galaxies directly with the spectroscopic studies. In this paper
 we adopt the
fundamental plane for ellipticals to estimate the central velocity 
dispersion and black hole mass for a number of AGNs with morphology 
parameters of their elliptical host galaxies obtained by the {\it Hubble Space 
Telescope} imaging observations.
The estimated black hole masses of 63 BL Lac objects, 10 radio galaxies, 
10 radio-loud quasars and 9 radio-quiet quasars are mostly in the range of $10^{7.5}
M_\odot$ to    $10^{9} M_\odot$. No significant difference in black hole
mass is found for high-frequency peaked BL Lacs and low-frequency peaked
BL Lacs, as well as  for radio galaxies and radio-loud quasars. 
The Eddington ratios of radio galaxies are
substantially smaller than those of quasars.
This
suggests that the different observational features of these radio-loud AGNs may be
mainly dominated by different accretion rate  rather than by
the black hole mass, which is in agreement with some evolutionary scenarios recently
proposed for radio-loud AGNs. 
Different from some previous claims, 
 we found that the derived mean black hole mass for radio-loud quasars is only 
slightly larger than that of radio-quiet quasars. Though the black hole mass 
distributions between radio-loud and radio-quiet quasars  are statistically 
different, their Eddington ratio distributions are probably from the same population.
 In addition, we noted that the relation
between black hole mass and host galaxy luminosity we obtained 
using the fundamental plane 
provides further arguments for a nonlinear scaling law between
supermassive  black hole and galactic bulge mass.

   }
\authorrunning{Wu et al.}
\titlerunning{Supermassive black hole masses of AGNs}

   \maketitle
%
%________________________________________________________________

\section{Introduction}

The masses of central black holes (M$_{BH}$) in about 40 nearby galaxies have 
been recently obtained using  the stellar and gas dynamics methods by the {\it Hubble 
Space Telescope} (HST) (for a recent review see Kormendy \& Gebhardt 2001). A tight 
correlation between black hole mass and bulge velocity
dispersion ($\sigma$) has been found for nearby galaxies (Gebhardt et al. 2000a; 
Ferrarese \& 
Merritt 2000). With the reverberation mapping technique (Netzer \& Peterson 1997), the
supermassive black hole (SMBH) masses of about 20 Seyfert 1 galaxies 
(Wandel, Peterson \& Malkan 1999;
Ho 1999) and 20 nearby quasars (Kaspi et al. 2000)
were estimated based on a virial assumption about the dynamics of 
the broad line region of Active Galactic Nuclei (AGNs). Interestingly, 
the black hole masses derived
for a few Seyfert galaxies with measured central velocity dispersions
of their host galaxies follow the same M$_{BH}$-$\sigma$ relation as
nearby galaxies (Gebhardt et al 2000b; Ferrarese et al. 2001). This 
indicates that the  M$_{BH}$-$\sigma$ relation is probably  universal
for both active and inactive galaxies. The small scatters of this
relation also imply that it may be more fundamental than the relation
between black hole mass and bulge luminosity (Magorrian et al. 1998). The
close correlation between black hole mass and bulge properties has important
implications to the formation and evolution of SMBHs and 
galaxies. 

On the other hand, the tight M$_{BH}$-$\sigma$ relation suggests an interesting
possibility to estimate the central black hole masses for galaxies using the
measured values of bulge velocity dispersions. This straightforward
method is particularly important
for  AGNs because the dynamical method can not be applied for the determinations of 
the black hole masses for most of them. For some AGNs, especially BL Lacertae objects, 
 the reverberation mapping technique can not be applied  because they 
have no or only very weak emission lines in their optical spectra.  However,
AGNs usually have very bright nuclear emission, which makes it very difficult
to measure their stellar velocity dispersions with the spectroscopic method.
So far, stellar velocity dispersions have been obtained only for some
nearby Seyfert galaxies (Nelson \& Whittle 1995; Ferrarese et al. 2001)
and recently for one nearby BL~Lac object Mrk~501 (Barth, Ho \& Sargent 2002). For most
AGNs, one has to look for other methods to determine the central velocity 
dispersions of their host galaxies in order to use the M$_{BH}$-$\sigma$ 
relation to estimate the SMBH masses.

Imaging studies on the host galaxies of AGNs with HST have 
clearly revealed that a lot of AGNs,
including almost all BL Lac objects, radio galaxies, radio-loud quasars, and
some radio-quiet quasars, have massive elliptical hosts (Urry et al. 2000;
McLure et al. 1999; Dunlop et al. 2002).  It is well known for  
ellipticals that three observables, the effective radius, 
the corresponding average surface
brightness and the central velocity dispersion, follow a surprisingly
tight linear relation (so called fundamental plane, see Djorgovski
\& Davis 1987; Dressler et al. 1987; Faber et al. 1989). 
Some subsequent studies have shown that  the elliptical hosts of radio 
galaxies follow the same fundamental plane
as normal ellipticals (Bettoni et al. 2001).
Because the fundamental plane is probably universal and  exists also for
elliptical hosts of AGNs, it is possible to estimate the central velocity dispersions
from the morphology parameters of the host galaxies (McLure \& Dunlop
2001). This provides
another possible way to derive the SMBH masses for AGNs which have
been obtained high quality images of their host galaxies.

In this paper we  adopt the fundamental plane to estimate the central
velocity dispersions and SMBH masses for some AGNs which have been imaged by HST. 
In section 2 we introduce
the fundamental plane for AGN elliptical hosts. The SMBH
 masses of these AGNs are derived in Section 3. In Section 4 
the physics nature of these AGNs are briefly discussed based on our results.

%__________________________________________________________________

\section{Fundamental plane of elliptical galaxies}

The fundamental plane of ellipticals has been extensively studied and
well established with the
ground based observations ( Djorgovski
\& Davis 1987; Dressler et al. 1987; Faber et al. 1989; Jorgensen, Frank \& Kjaergaad 1996).
Such a plane has been shown to be close to the plane defining the viral equilibrium if
a rigorous homology among galaxies is assumed (Faber et al. 1989).
Imaging studies on the host elliptical galaxies of low redshift radio galaxies found
a similar fundamental plane as for inactive elliptical galaxies
(Bettoni et al. 2001), with radio galaxies representing
the brightest end of the population of early type galaxies. 
This also implies that the global properties of early-type galaxies are not influenced
by the gas accretion process around the central black hole.
It is therefore quite likely that not only radio galaxies but also other AGNs with  
elliptical host galaxies  follow the similar
fundamental plane as  normal ellipticals.

Using the observational data of about 300 normal ellipticals and radio galaxies,
Bettoni et al. (2001) found that the fundamental plane can be robustly described as 
$$
\log R_e = (1.27\pm0.04) \log \sigma + (0.326\pm0.007) <\mu_e>_R 
$$
\begin{equation}
~~~~~~~~~~~~~~~~~~~~~~~~~~~~~~~~~~~~~~- 8.56\pm0.06,
\end{equation}
where $R_e$ is the effective radius in $kpc$, $\sigma$ is the central velocity
dispersion in $km s^{-1}$, and $<\mu_e>_R$ is the average surface brightness in R-band.
If we assume that all AGNs with elliptical hosts follow this fundamental plane, we can
estimate their central velocity dispersions based on the morphology parameters,
$R_e$ and $<\mu_e>_R$, which can be derived from high quality imaging studies of
their host galaxies.

\section{Black hole masses of AGNs}

In this section we will adopt the fundamental plane and the M$_{BH}$-$\sigma$ relation
to estimate the SMBH masses of some BL Lac objects, radio galaxies and quasars that
have been imaged by the HST recently. The higher spatial resolution of HST can provide
high quality images of the host galaxies of AGNs, which enables us to reliably derive  
the morphology parameters.

\subsection{BL Lac objects}
 
The BL Lac snapshot survey using the HST WFPC2 camera has obtained images for 110
BL Lac objects in a well selected sample (Scarpa et al. 1999). By fitting the surface 
brightness profiles using de Vaucouleurs model, Urry et al. 
(2001) has obtained
host galaxy parameters for 72 BL Lac objects. They showed that these detected hosts 
are very luminous, round galaxies with a median absolute magnitude of $<M_R>=-23.7$ mag
and a median effective radius of $<R_e>=8.5$ kpc (we used
$ H_0=50 \rm{kms^{-1}Mpc^{-1}}$ and $q_0=0$ throughout the paper). Among these BL Lacs, 63 objects
have measured redshifts. 51 of them are classified as high-frequency peaked BL Lacs (HBL)
and 12 of them as  low-frequency peaked BL Lacs (LBL). The morphology parameters of their
elliptical hosts, including the angular effective radius $r_e$, corresponding surface brightness 
$\mu_e$ and
absolute R magnitude $M_R$,
 have been reported in Urry et al. (2001). It has been shown that the $\mu_e$-$r_e$ relation
for the elliptical hosts of these BL Lac objects is all most the same as normal elliptical
galaxies. This strongly suggests the the fundamental plane of normal ellipticals exists also
for the host galaxies of BL Lac objects. Urry et al. (2001) also mentioned that there are no
systematic differences in the host galaxies of
HBLs and LBLs.

%__________________________________________________________________
%__________________________________________________ two column table
%\begin{center}

 \tabcolsep 2.5mm
  \begin{table*}
      \caption[]{Sample of BL Lac objects} %%
\scriptsize
         \label{table1}
         \begin{tabular}{lccccccccc}
            \hline
            \noalign{\smallskip}       
Name &           z    &      R &    $M_R$ &    $r_e$   &    $R_e$ &    $<\mu_e>$    
& $ \log~ \sigma $ &      $ \log~ M_{BH}^{MF01} $  & $ \log~ M_{BH}^{G00} $ \\
 & & (mag) & (mag) & (arcsec) & (kpc) & ($ R~ mag~ arcsec^{-2} $) & ($km~ s^{-1}$) & ($ M_\odot $)  & $ (M_\odot $) \\   
\hline
HBL & & & & & & & & &\\
\hline
0122+0908   &      0.339    &     17.5    &   -23.75     &    1.05     &    6.75  &       19.6    &    2.36   &     8.40      &   8.31\\
0145+1388     &    0.124     &    16.5      & -22.74   &      1.75    &     5.31    &    19.71     &   2.25     &   7.88     &   7.89\\
0158+0018     &    0.229    &    17.43    &   -23.05   &       1.9     &    9.34      &  20.82      &  2.15     &   7.44      &  7.54\\
0229+2008    &     0.139   &     14.87   &    -24.61      &   3.25     &   10.83    &    19.42   &     2.56   &      9.38      &  9.08\\
0257+3428    &     0.247  &      16.58   &    -24.05     &    1.75     &    9.08    &    19.79      &  2.41     &   8.64     &   8.50\\
0317+1838     &     0.19     &   16.39   &    -23.71   &      3.25      &  13.89   &     20.95    &    2.26     &   7.93   &     7.93\\
0331-3628     &    0.308   &     16.74    &   -24.33       &   3.1   &     18.73    &    21.19  &      2.30    &    8.12      &  8.08\\
0347-1218      &   0.188     &   16.89  &      -23.2     &    1.25     &     5.3   &     19.37   &     2.33     &   8.29    &    8.22\\
0350-3718     &    0.165   &     16.47     &  -23.35   &       1.7    &     6.51      &  19.62    &    2.34     &   8.32      &  8.24\\
0414+0098     &    0.287     &   16.07   &    -24.85    &      4.7    &    27.09     &   21.43    &    2.36    &     8.43    &     8.33\\
0419+1948     &    0.512    &    18.03    &   -24.01     &     0.4     &    3.27     &   18.04    &    2.51     &   9.12   &     8.88\\
0502+6758    &     0.314   &     17.22     &  -23.88    &      0.6       &  3.67    &     18.1      &  2.53     &   9.23       & 8.96\\
0506-0398     &    0.304   &     17.21     &  -23.78     &     1.6        & 9.57    &    20.23      &   2.32    &    8.20    &    8.14\\
0525+7138      &   0.249  &      16.17     &  -24.48       &  1.98    &    10.34       & 19.65     &   2.49    &    9.03     &    8.81\\
0548-3228    &     0.069    &    14.31   &    -23.71    &     7.05      &  12.81   &     20.54      &  2.33      &  8.29    &    8.22\\
0607+7108     &    0.267   &     16.46   &    -24.34     &     2.4     &   13.16  &      20.35     &   2.39    &    8.56  &      8.43\\
0706+5918     &    0.125     &   15.22     &  -24.03    &     3.05    &     9.31     &   19.64     &   2.46     &   8.87      &   8.68\\
0737+7448     &    0.315    &    16.79     &  -24.32      &    2.1      &  12.88    &    20.39     &    2.38   &     8.48    &    8.37\\
0806+5248     &    0.138    &    15.93     &  -23.53     &    1.45     &     4.8      &  18.73     &   2.46     &   8.90     &   8.70\\
0922+7498      &   0.638     &   17.82     &  -24.64      &   0.85     &    7.76    &    19.46     &   2.44    &    8.79     &    8.62\\
0927+5008     &    0.188   &     16.95   &    -23.14       &     2     &    8.48      &  20.45     &   2.22   &     7.74    &    7.78\\
0958+2108     &    0.344    &    17.65  &     -23.62    &     0.82      &   5.32    &    19.21     &    2.38     &   8.48     &   8.37\\
1011+4968        &   0.2    &    16.61     &   -23.6       &   1.8       &     8    &    19.89     &   2.34       & 8.32      &   8.25\\
1028+5118      &   0.361      &  17.31  &     -24.07  &        1.8      &  12.05    &    20.58       & 2.30  &      8.14  &      8.10\\
1104+3848      &   0.031  &       13.1    &   -23.21   &      3.95     &     3.4    &    18.08      &  2.51     &   9.13      &  8.88\\
1133+1618       &   0.46     &   18.08     &  -23.76   &      1.55      &  11.97    &    21.03    &    2.19   &     7.59   &      7.67\\
1136+7048     &    0.045   &     14.22      &  -22.9      &    3.1       &   3.8   &     18.68      &  2.40     &   8.59     &    8.46\\
1207+3948     &    0.615     &   17.98     &   -24.4     &     1.2    &    10.76   &     20.37      &  2.32     &   8.22    &    8.16\\
1212+0788      &   0.136     &    15.5    &   -23.93      &    3.4      &  11.13    &    20.16      &   2.39     &   8.53   &     8.41\\
1215+3038     &     0.13    &    15.52    &    -23.8      &   8.35      &  26.34    &    22.13      &  2.17   &     7.53   &      7.62\\
1218+3048   &      0.182     &   16.46   &    -23.56       &   2.6   &     10.75       & 20.53     &   2.28    &    8.02   &     8.01\\
1221+2458     &    0.218      &  17.89    &   -22.49     &    1.25      &   5.93   &     20.37      &   2.12      &  7.25      &  7.39\\
1229+6438    &     0.164     &   15.74    &   -24.07         &   2    &     7.62    &    19.24     &   2.49     &   9.03  &      8.80\\
1248-2968      &    0.37  &      17.28  &     -24.14    &      1.1    &     7.47      &  19.48      &  2.42   &     8.71     &   8.55\\
1255+2448    &     0.141   &     16.18   &    -23.32     &     2.5       &  8.43   &     20.17     &   2.29   &     8.07   &     8.04\\
1407+5958    &     0.495   &      17.2    &   -24.78    &     1.75      &  14.06   &     20.41     &   2.40    &    8.60    &    8.47\\
1426+4288     &    0.129    &    15.64   &    -23.68      &   2.25   &      7.05      &  19.39      &   2.43     &   8.72      &  8.56\\
1440+1228      &   0.162   &     16.11     &  -23.68     &     3.9    &    14.71     &   21.06      &  2.25  &      7.88   &     7.89\\
1458+2248       &  0.235     &   16.84  &     -23.69     &     3.2    &    16.03   &     21.36     &   2.20      &  7.66      &   7.7\\
1514-2418     &    0.049    &    13.76    &   -23.54   &       3.7      &   4.91   &     18.59      &  2.51   &     9.10  &      8.86\\
1534+0148      &   0.312    &    16.89    &   -24.21     &       2      &  12.19    &    20.39     &   2.36    &    8.39     &   8.30\\
1704+6048     &     0.28      &  17.69      &  -23.2   &      0.85      &   4.82    &    19.34    &    2.31   &     8.17      &  8.12\\
1728+5028      &   0.055   &     15.15     &  -22.38   &      3.15      &   4.65    &    19.64      &  2.22       & 7.75     &   7.79\\
1757+7038     &    0.407    &    17.97   &    -23.63     &    0.85       &  6.12     &   19.61      &  2.32     &    8.23      &  8.17\\
1853+6718       &  0.212  &      17.17    &   -23.16     &     1.5    &     6.96     &   20.05      &  2.25    &    7.90  &      7.91\\
1959+6508     &    0.048     &   14.13     &  -23.12    &      5.1    &     6.64     &   19.66      &  2.34    &    8.30    &    8.22\\
2005-4898      &   0.071     &   14.22       &-23.86   &      5.65      &  10.53    &    19.98      &  2.41    &    8.66     &   8.51\\
2143+0708     &    0.237     &   16.87  &     -23.68      &    2.1   &     10.58   &     20.47      &  2.29   &     8.06   &     8.04\\
2326+1748     &    0.213     &   16.67   &    -23.67        &  1.8    &     8.39    &    19.94    &    2.34      &  8.33   &     8.25\\
2344+5148    &     0.044    &    12.89    &   -24.19      &   5.93     &    7.12     &   18.75     &   2.59   &     9.52    &    9.19\\
2356-3098    &     0.165     &    16.6    &   -23.22      &   1.85      &   7.08    &    19.93      &  2.29       & 8.07    &    8.05\\
\hline
LBL & & & & & & & & & \\
\hline
0521-3658     &     0.055    &     14.23     &    -23.3      &     2.8    &      4.14   &      18.46    &    2.48    &    8.98     &   8.77\\
0828+4938      &   0.548     &   18.03   &    -24.14     &    0.65     &    5.51  &      19.09      &  2.42   &      8.69     &   8.53\\
0829+0468     &     0.18      &  16.18   &    -23.82     &     4.3     &   17.63   &     21.34     &   2.24   &     7.83   &     7.86\\
1418+5468      &   0.152   &     15.56    &   -24.09    &     3.65   &     13.08  &      20.37      &  2.39     &   8.53      &  8.41\\
1538+1498       &  0.605       &  17.7   &    -24.64      &    2.5     &   22.25   &     21.69     &   2.23     &   7.79    &    7.82\\
1749+0968       &   0.32   &     17.55   &     -23.6     &       3      &  18.59    &    21.93     &   2.11   &     7.21     &   7.36\\
1807+6988      &   0.051     &   13.43   &    -23.95       &   2.1   &      2.89      &  17.04     &   2.72      &  10.14    &    9.68\\
1823+5688    &     0.664   &     17.46   &    -25.07       &   0.6   &      5.57      &  18.35    &    2.61     &   9.60       & 9.26\\
2007+7778     &    0.342     &   17.38      & -23.89       &   3.3   &     21.35   &     21.97      &  2.14    &     7.39    &    7.50\\
2200+4208      &   0.069      &  14.41     &  -23.61  &        4.8      &   8.72  &      19.81      &  2.39    &    8.56   &     8.43\\
2201+0448    &     0.027  &      13.44     &  -22.57 &        6.78       &  5.12  &      19.59     &   2.26     &   7.96      &  7.95\\
2254+0748        &  0.19   &     15.69   &    -24.41   &       4.9     &   20.94  &      21.14     &   2.35    &    8.36    &    8.27\\

         \noalign{\smallskip}
            \hline
         \end{tabular}

%\begin{list}{}{}
%\item{}
%$^*$Notes:  
%\end{list}
   \end{table*}

In Table 1 we listed 63 BL Lac objects, together with their redshifts, R-band apparent and absolute
magnitudes, effective radius and average surface brightness of their host galaxies. The values
for redshifts, R-band absolute magnitudes, and  effective radii are taken from
Urry et al. (2001). The R-band apparent magnitudes have been corrected for
 Galactic extinctions,
cosmological dimming and K-corrections. The average surface 
brightness was derived from the formula:
\begin{equation}
<\mu_e>_R=R +5\log r_e +2.5\log (2\pi)
\end{equation}
where the effective radius $r_e$ is in unit of arcsecends. We can then estimate the central
velocity dispersions for the hosts of BL Lac objects using the fundamental plane (Eq. (1)).
The SMBH masses of these objects can be derived by the $M_{BH}$-$\sigma$ relation. Such a
relation has been given as
\begin{equation}
M_{BH}=1.3 \times 10^8M_\odot (\sigma/200 kms^{-1})^{4.72}
\end{equation}
by Merritt \& Ferrarese (2001, hereafter MF01) and as
\begin{equation}
M_{BH}=1.2 \times 10^8M_\odot (\sigma/200 kms^{-1})^{3.75}
\end{equation}
by Gebhardt et al. (2000a, hearafter G00). 
Note that The derived central velocity dispersion using Eq. (1) is in 
 an aperture of diameter of $1.19h^{-1}$kpc ($h=0.5$ in this paper), while the 
luminosity-weighted average
velocity dispersion in G00 relation is in the half-light radius ($r_e$) and the 'central'
velocity dispersion in MF01 relation in  an aperture of radius $r_e/8$. There is a 
systematic difference among these velocity dispersions. However, as demonstrated in
G00, MF01 and Barth et al (2002), such a difference is remarkably less significant when
$r\le r_e$. In fact,
if we follow the aperture correction method suggested by Jorgensen, Frank, \& Kjaergaard 
(1995), we can estimate the difference of these velocity dispersions to be as small as
several percents. Therefore, we ignore such a difference in our present work.  
In Table 1 we give the derived central
velocity dispersions and black hole masses using the above relations for 63 BL Lacs.

%
%                                                One column figure
%----------------------------------------------------------- S_vib
   \begin{figure}
   \centering
   \includegraphics[width=8.7cm, height=11cm]
{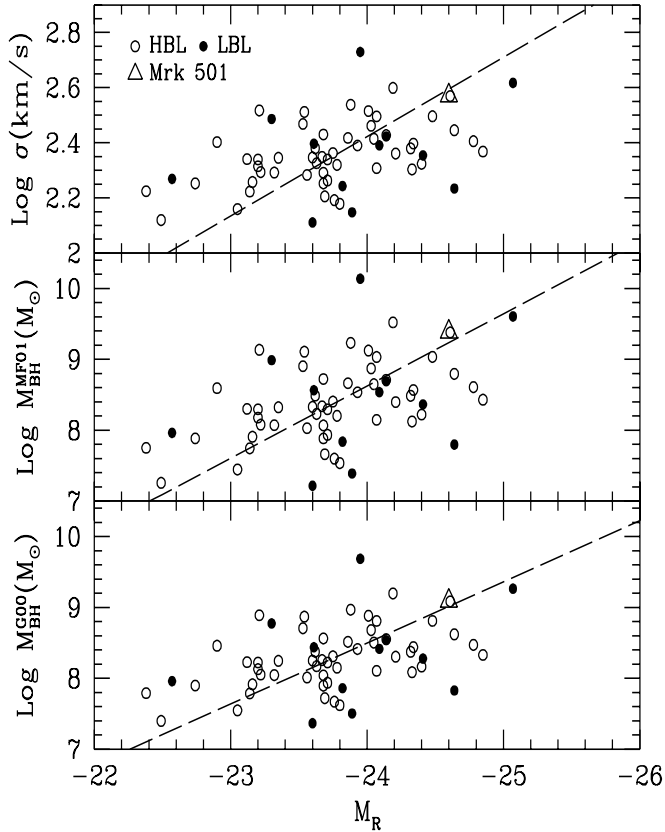}
      \caption{The derived central velocity dispersion, black hole mass (using both MF01 and G00
relations) for BL Lac objects against
             the R-band absolute magnitude of the host galaxies. The open and solid circles 
correspond to HBLs and LBLs. The dashed line shows the OLS bisector fit to each relation. The open
triangle represents the data for Mrk~501.}
         \label{Fig1}
   \end{figure}
%
%______________________________________________________________

%
%                                                One column figure
%----------------------------------------------------------- S_vib
   \begin{figure}
   \centering
   \includegraphics[width=8.7cm]
{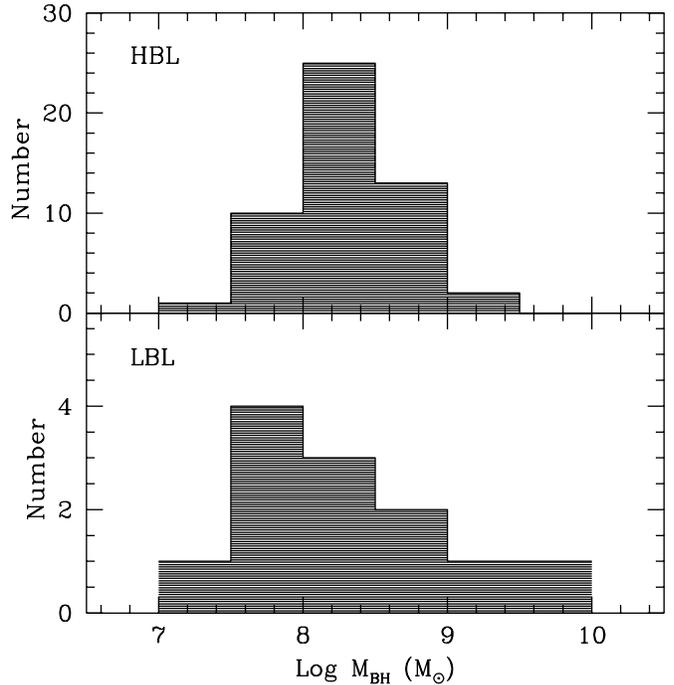}
      \caption{Histogram of the derived black hole mass distribution of HBLs and LBLs using the 
G00 relation.}
         \label{Fig2}
   \end{figure}
%
%______________________________________________________________

Figure 1 shows the relation of derived central velocity dispersions and black hole masses
with R-band absolute magnitudes of the host galaxies of BL Lac objects. Figure 2 shows the
histograms of the derived black hole masses (according to the G00 relation) for HBLs and LBLs.
It is clear that there is no significant difference in the central velocity dispersions and
the black hole masses between HBLs and LBLs. The average black hole masses of both HBLs and LBLs
are around $2\times 10^8 M_\odot$. Most of BL Lacs have black hole masses in the
range of $10^{7.5}M_\odot$ to  $10^{9.5}M_\odot$. 
A T-test gives a significance of
81\% that the distributions of SMBH masses of HBLs and LBLs are from the population
with the same true variance. 
Considering the substantial uncertainties in
deriving the R-band absolute magnitude and central velocity dispersions of host galaxies,  we adopt 
the ordinary least square (OLR) bisector method (Isobe et al. 1990) to fit
 the relations shown in Figure 1, which gives: 
$\log \sigma = -4.49\pm1.30-(0.29\pm0.05)M_R$,
$\log M_{BH}^{MF01}=-15.83\pm2.08-(1.01\pm0.08)M_R$, and
$\log M_{BH}^{G00}=-12.18\pm1.95-(0.86\pm0.08)M_R$.
We noted that 
the $ M_{BH}$-bulge luminosity relation we  obtained here is consistent with that found previously
for normal galaxies and AGNs (Laor 2001; Wu \& Han 2001). 

Recently Barth et al. (2002) measured the calcium triplet lines in the spectra of a nearby
bright BL Lac object Mrk~501 taken with the Palomar Hale 200-inch telescope and derived
the stellar velocity dispersion of the host galaxy as $372\pm 18 \rm{kms^{-1}}$. The R-band
absolute magnitude of the host galaxy has been estimated to be -24.6 (Nilsson et al. 1999).
From Figure 1 we can see the observational data of Mrk~501 are well agreement with
the relation that we found for other BL Lac objects. Therefore we think it is quite possible that
Mrk~501 hosts a supermassive black hole with mass of about $(1\sim3) \times 10^9M_\odot$.
We note that this is much larger than the maximum primary black hole mass ($10^8M_\odot$)
required for the binary black hole model of Mrk~501 (Rieger \& Mannheim 2000) and
the mass ($\sim 10^7M_\odot$) estimated from the $\gamma$-ray variability timescale (Fan, Xie \& Bacon 1999).

We note that there are significant uncertainties in our derived velocity dispersions
and SMBH masses, which are caused by the systematic errors of fundamental plane relation
(Eq. (1)), 
the measurement uncertainties of $<\mu_e>_R$ and $R_e$, and the scatters of the 
$M_{BH}$-$\sigma$ relation. Considering the typical uncertainties of $<\mu_e>_R$ ( about 0.5 
$R~mag~arcsec^{-2}$) and  $R_e$ ( about 1 $kpc$), Eq. (1) gives an uncertainty
$\Delta \sigma/\sigma $ as large as of about 50\%. This leads to an uncertainty 
 $\Delta M_{BH}/M_{BH}$  of about 2. Such a significant uncertainty of the derived SMBH 
masses should be kept in mind when these values are adopted in any  
correlation studies.

\subsection{Radio galaxies, radio-loud and radio-quiet quasars}

%__________________________________________________________________
%__________________________________________________ two column table
%\begin{center}

 \tabcolsep 2.5mm

  \begin{table*}
	\scriptsize
      \caption[]{Sample of radio galaxies, radio-loud and radio quiet quasars} %%
         \label{table2}
         \begin{tabular}{lccccccccc}

            \hline
            \noalign{\smallskip}         
Name &           z    &      R &    $M_R$ &    $r_e$   &    $R_e$ &    $<\mu_e>$    
& $ \log~ \sigma $ &      $ \log~ M_{BH}^{MF01} $  & $ \log~ M_{BH}^{G00} $ \\
 & & (mag) & (mag) & (arcsec) & (kpc) & ($ R~ mag~ arcsec^{-2} $) & ($km~ s^{-1}$) & ($ M_\odot $)  & $ (M_\odot $) \\   
\hline
RG & & & & & & & & &\\
\hline

0230-027    &     0.239   &     16.84  &     -23.67   &     1.61    &      7.7     &   19.88     &   2.33     &   8.27   &     8.20\\
0307+169   &      0.256   &     16.24    &   -24.09   &     1.88    &      9.4    &    19.61   &     2.47    &    8.91     &   8.71\\
0345+337    &     0.244     &   16.85    &   -23.17    &     2.71   &      13.1     &   21.01     &   2.22     &   7.75     &   7.79\\
0917+459     &    0.174    &    15.63   &     -24.29     &   5.73      &   21.9    &    21.42   &     2.29     &     8.10      &  8.06\\
0958+291      &   0.185   &     16.59    &   -23.45      &  2.12    &      8.5     &   20.22   &     2.28      &  8.01  &      8.00\\
1215-033   &      0.184  &      16.55  &     -22.94      &  2.13    &      8.5    &    20.19     &   2.28       & 8.05     &   8.03\\
1215+013    &     0.118   &     16.14      & -23.35   &     1.66   &       4.7     &   19.25   &     2.329     &   8.24   &     8.18\\
1330+022     &    0.215    &     16.5   &     -23.81    &    3.53     &    15.7    &    21.24    &    2.22      &  7.77  &      7.80\\
1342-016     &    0.167    &    15.06      & -24.67   &     6.29       &  23.3  &      21.05   &     2.41       & 8.64   &     8.49\\
2141+279      &   0.215     &   15.93   &    -24.20   &     5.58      &   24.8 &       21.66     &   2.27      &  8.00    &    7.99\\

\hline
RLQ & & & & & & & & &\\
\hline
0137+012      &   0.258    &    16.49    &   -24.17   &     2.83       &  14.2    &    20.75   &     2.32    &    8.21    &    8.15\\
0736+017      &   0.191    &    16.11   &    -23.68     &   3.25       &  13.3    &    20.67  &      2.31      &  8.19   &     8.14\\
1004+130     &     0.24    &    16.21    &    -24.22      &  1.71  &        8.2    &    19.38    &    2.48       & 8.98   &     8.76\\
1020-103       &  0.197    &    16.59  &     -23.46       &   1.70     &     7.1    &    19.74     &   2.34     &   8.31    &    8.24\\
1217+023    &      0.24    &    16.66     &  -23.83      &  2.32     &    11.1 &       20.48    &    2.30     &   8.13  &      8.09\\
2135-147       &    0.2    &    16.57    &   -23.47    &    2.74    &     11.6   &     20.76   &      2.25      &  7.87    &    7.88\\
2141+175      &   0.213    &    16.47    &   -23.54 &       1.85   &       8.2      &  19.81   &     2.37      &  8.46     &   8.35\\
2247+140       &  0.237   &     16.51       & -23.92    &    2.84      &   13.5   &     20.78   &     2.29       & 8.09     &   8.06\\
2349-014     &    0.173    &    15.39      & -24.41       & 5.05   &      19.2    &    20.91  &      2.38      &  8.50     &    8.39\\
2355-082     &     0.21  &      16.48   &    -23.73    &    2.38     &    10.4   &     20.35  &      2.31      &  8.18   &     8.13\\
\hline
RQQ & & & & & & & & &\\
\hline
0054+144    &     0.171   &     16.05    &   -23.70   &      2.76      &   10.4     &   20.25      &  2.34      &  8.31     &   8.23\\
0204+292     &    0.109 &       15.45     &   -23.36    &     3.33     &     8.8    &    20.05   &     2.33     &   8.27   &      8.21\\
0244+194     &    0.176    &     16.8   &    -22.36   &     2.41       &   9.3     &   20.71    &    2.18      &  7.57      &  7.65\\
0923+201       &   0.19     &   16.61   &    -23.35   &     2.01    &      8.2   &     20.13      &  2.29      &  8.07  &      8.04\\
0953+414      &   0.239    &    17.58     &  -22.98  &      1.59       &   7.6    &    20.59    &    2.14      &  7.39   &     7.50\\
1012+008      &   0.185   &     16.05     &  -23.87    &    7.18     &    28.7    &    22.32     &   2.15      &  7.43   &     7.54\\
1549+203   &       0.25  &      18.15   &    -22.38    &    1.01      &      5     &   20.19    &    2.10       & 7.20   &     7.35\\
1635+119    &     0.146  &      16.28    &   -23.13   &     2.27     &     7.6      &  20.06    &    2.28      &  8.03    &    8.01\\
2215-037     &    0.241   &     16.64    &   -23.64   &     1.39   &       6.7     &   19.36    &    2.42     &   8.68    &    8.53\\

         \noalign{\smallskip}
            \hline
         \end{tabular}

%\begin{list}{}{}
%\item{}
%$^*$Notes:  
%\end{list}
   \end{table*}

A deep HST imaging study of the host galaxies of a sample of 10 radio galaxies (RGs), 10 radio-loud
quasars (RLQs) and 13 radio-quiet quasars (RQQs) has been  performed recently (McLure et al. 1999; 
Dunlop et al. 2002).  It has been found that the hosts of both radio-loud AGNs and bright 
radio-quiet AGNs are virtually all massive ellipticals. The basic properties of these host 
galaxies are indistinguishable from those of normal ellipticals. Therefore, it is quite possible
that the host galaxies of these low redshift AGNs also follow the same fundamental plane 
as normal ellipticals.
Using the same approach as we did for BL Lac objects, we can also derive the central
velocity dispersions and black hole masses for this sample of AGNs based on the morphology
parameters of their host galaxies.

Table 2 listed the name, redshift, R-band apparent magnitude and absolute magnitude, effective 
radius and average surface brightness of 10 RGs, 10 RLQs and 9 RQQs with elliptical hosts 
(Dunlop et al. 2002). The R-band apparent magnitudes of the host galaxies 
have been corrected for
the Galactic extinctions (taken from NED\footnote{http://nedwww.ipac.caltech.edu}), cosmological dimming and
K-corrections (assuming spectra index of $\alpha=1.5$, Dunlop et al. 2002).
The angular size of effective radius was derived using Eq. (9.94) in Peterson (1997) from
$R_e$ listed in Dunlop et al. (2002) (where they adopted $q_0=0.5$):
\begin{equation}
r_e('') =\frac{0.0688h_0q_0^2(1+z)^2}{zq_0+(q_0-1)(-1+\sqrt{2q_0z+1})} R_e(\rm{kpc}).
\end{equation}
%Here we used $h_0=H_0/100\rm{kms^{-1}Mpc^{-1}}=0.5$. 
The 
R-band absolute magnitudes of the host galaxies were also calibrated to the case of
$q_0=0$. The average surface brightness was calculated using Eq. (2) for each objects.
With these parameters, the central velocity dispersions of the host galaxies of 
29 AGNs can be estimated using Eq. (1), and their SMBH masses can be derived using
Eq. (3) and (4). The results are  listed in Table 2.

%
%                                                One column figure
%----------------------------------------------------------- S_vib
   \begin{figure}
   \centering
   \includegraphics[width=8.7cm,height=11.cm]
{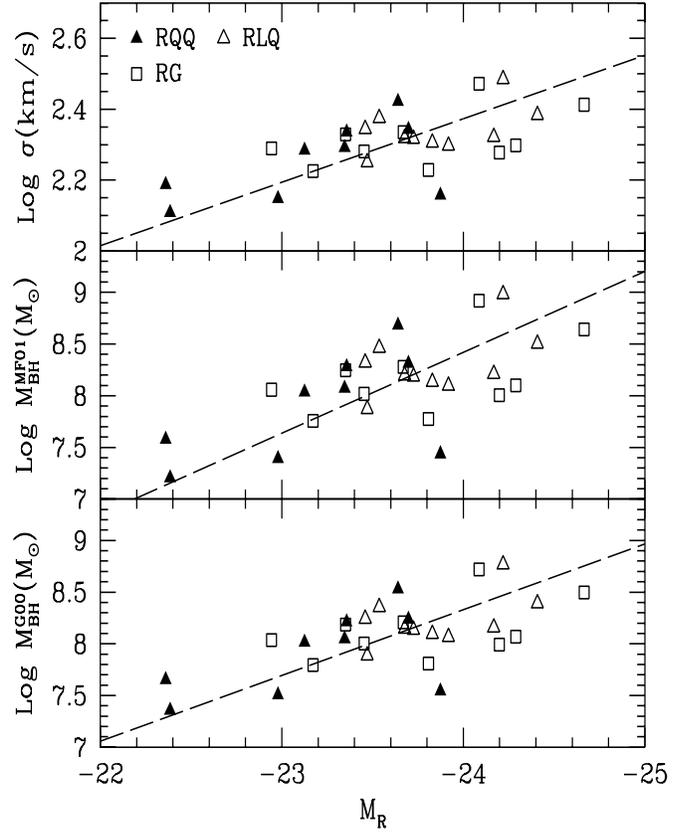}
      \caption{The derived central velocity dispersion, black hole mass (using both MF01 and G00
relations) for AGNs against
             the R-band absolute magnitude of the host galaxies. The solid and open 
triangles represent radio-quiet and radio-loud quasars. The open squares represent radio
galaxies. The dashed line shows the OLS bisector fit to each relation.}
         \label{Fig3}
   \end{figure}
%
%______________________________________________________________

%
%                                                One column figure
%----------------------------------------------------------- S_vib
   \begin{figure}
   \centering
   \includegraphics[width=8.7cm]
{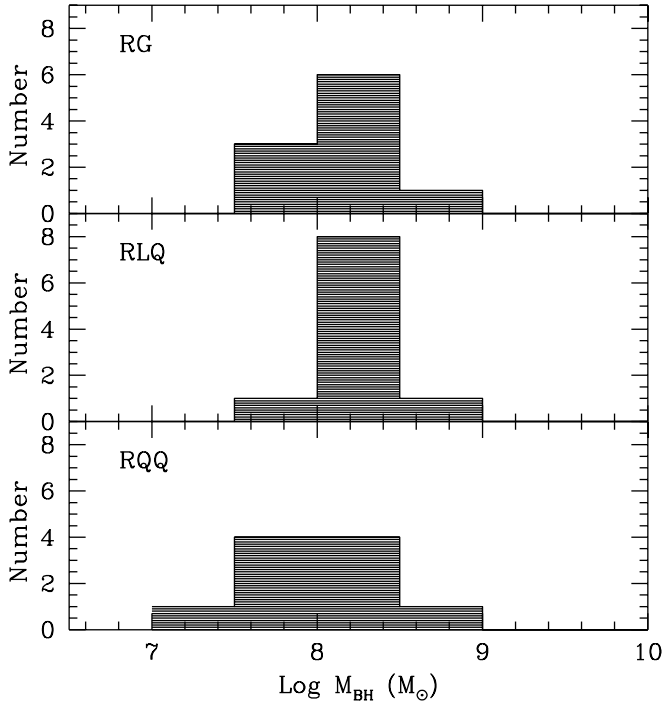}
      \caption{Histogram of the derived black hole mass distribution of radio galaxies, 
radio-loud and radio-quiet quasars using the G00 relation.}
         \label{Fig4}
   \end{figure}
%
%______________________________________________________________

Figure 3 shows the relation of derived central velocity dispersions and black hole masses
with R-band absolute magnitudes of the host galaxies for 10 RGs, 10 RLQs and 9 RQQs. Figure 4
 shows the
histograms of the derived black hole masses (according to the G00 relation) for these AGNs.
It is clear that there is no significant differences in the central velocity dispersions and
SMBH masses among RGs, RLQs and RQQs. The average SMBH masses of RGs, RLQs and 
RQQs are $10^{8.13}M_\odot$, $10^{8.22}M_\odot$and $10^{7.90}M_\odot$ respectively.
Most of these AGNs have black hole masses in the
range of $10^{7.5}M_\odot$ to  $10^{9}M_\odot$. Our results indicate that there is no
difference in SMBH masses of BL Lacs, RGs and RLQs. A T-test shows a possibility of
44\% that the distributions of SMBH masses of RGs and RLQs are from the same population.
Different from some previous claims that
RLQs have more massive SMBHs than RQQs
(Laor 2000), our result shows that there is only weak  difference in our derived
SMBH masses for RQQs and RLQs in this sample. The mean SMBH mass of 9 RQQs
is smaller by only a factor of two than that of 10 RLQs. 
However, a T-test gives a possibility of only 4.5\% that the two distributions are
from the same population. Although it may indicate some statistical differences
between RLQs and RQQs (see also Dunlop et al. 2002), a more definitive conclusion about such
a difference can be reached only with  with larger and more 
complete samples of quasars. 
The OLR bisector fits
of the relations shown in Figure 3 give: 
$\log \sigma = -1.92\pm0.71-(0.18\pm0.03)M_R$,
$\log M_{BH}^{MF01}=-10.39\pm2.35-(0.78\pm0.10)M_R$, and
$\log M_{BH}^{G00}=-6.93\pm2.02-(0.64\pm0.08)M_R$.
These relations are slightly flatter than those we obtained for 63 BL Lac objects.
This may be caused by the smaller sample of 29 AGNs or several offset LBLs shown in Figure 1. 

We noted that the SMBH masses of 10 RLQs and 7 RQQs in our AGN sample have been estimated
in McLure \&Dunlop (2001) based on the $H\beta$ emission line width measurements and an 
empirical relation between broad line region size and optical luminosity (Kaspi et al. 2000).
By re-calculating the SMBH masses assuming the characteristic velocity in the broad line 
region of AGN can be estimated the observed FWHMs of $H\beta$ lines by 
$V_{BLR}=\frac{\sqrt{3}}{2}FWHM_{H\beta}$ (Wandel et al. 1999), in Figure
5 we plotted the comparison of the SMBH masses
derived by using the fundamental plane  with those obtained by $H\beta$ line study.  
The agreement is not  bad but
on average the SMBH masses estimated based on $H\beta$ line study are  
larger
than those obtained by us by a factor of two. If the relation between
$V_{BLR}$ and $FWHM_{H\beta}$ was assumed to be 
$V_{BLR}=1.5 \times FWHM_{H\beta}$ as did by McLure \& Dunlop (2001), the difference
between the SMBH masses obtained by  $H\beta$ line study will be
larger than our estimations by approximately a factor of 5. 
It is unclear whether such differences are due to our assumption of the fundamental
plane for AGN host galaxies or the overestimations of the broad line region sizes of AGNs
according to the empirical relation between the broad line region size and 
the optical luminosity (Kaspi \etal 2000). We note that the latter relation is much more scattered
than the previous one. Another advantage of our method is that it is model independent, while
the method  based on $H\beta$ line study sensitively depends on the assumptions
of the dynamics and geometry of broad line regions of AGNs (Krolik 2001). Considering
that both methods may have uncertainties of as larger as a factor of a few, the difference
in SMBH estimations shown in Figure 5 is not unexpected.

%
%                                                One column figure
%----------------------------------------------------------- S_vib
   \begin{figure}
   \centering
   \includegraphics[width=8.7cm,height=10.5cm]
{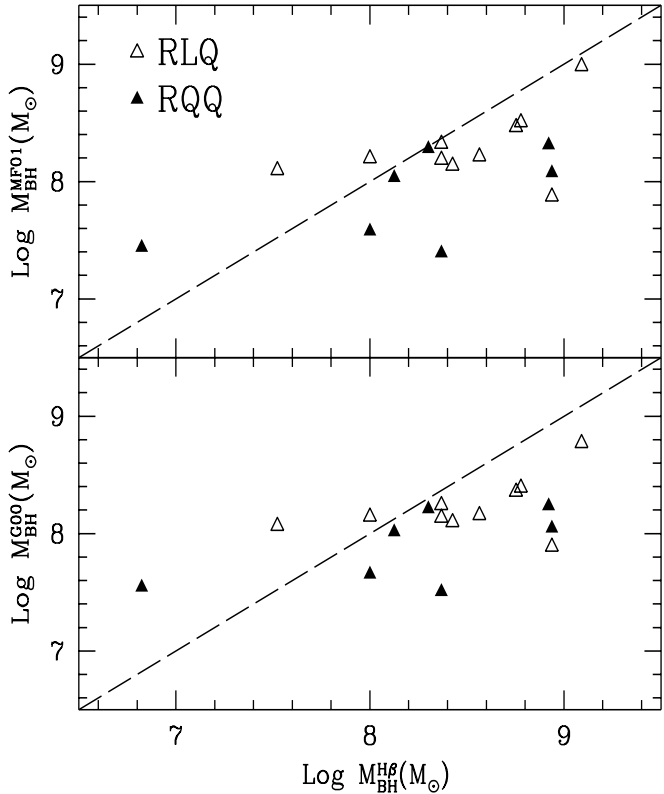}
      \caption{Comparison of the estimated SMBH masses of quasars by MF01 and G00 relations with those
derived from the $H\beta$ line study. The dashed line shows the one-to-one correspondence.}
         \label{Fig5}
   \end{figure}
%
%______________________________________________________________

%
%                                                One column figure
%----------------------------------------------------------- S_vib
   \begin{figure}
   \centering
   \includegraphics[width=8.7cm]%,height=10.5cm]
{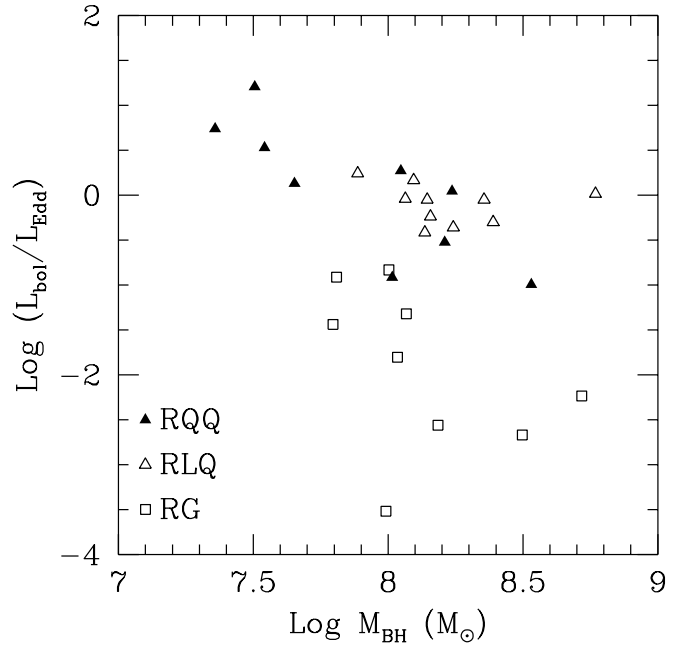}
      \caption{Comparisons of Eddington ratios of radio-loud quasars, radio-quiet quasars
and radio galaxies.}
         \label{Fig6}
   \end{figure}
%
%______________________________________________________________

Using the derived SMBH masses, we can estimate the Eddington ratio (defined as the
ratio of bolometric luminosity and Eddington luminosity) of the source in our sample 
of RGs, RLQs and RQQs. We adopted the assumptions of $L_{bol}\simeq 10 \lambda L_{5100\AA}$
(Kaspi et al. 2000) and $f_\nu \propto \nu^{-0.2}$ (Dunlop et al. 2002) to convert the 
R-band luminosity to the bolometric luminosity for the nuclear component of AGNs. Figure
6 shows the distributions of Eddington ratios of 9 RGs, 10 RLQs and 9 RQQs. It is clear
that the Eddington ratios of RGs are systematically smaller than those of RLQs and RQQs
by two orders, while there is less significant difference in Eddington ratios for RLQs
and RQQs. Our result is  qualitatively consistent with that 
obtained by Ho (2002) who recently suggested that the strongly active AGNs have larger
Eddington ratios. Figure 6 shows that both RLQs and RQQs have the bolometric luminosity 
comparable to the Eddington
luminosity. A T-test also shows a significance of 56\% that the Eddington ratios of RLQs
and RQQs are from the same population.   Therefore, our results indicate that the SMBH
masses of RLQs may be slightly larger than those of RQQs, their Eddington ratios may not be
significantly different. However, we must noted that these results were obtained with
a small sample of radio-loud and radio-quiet quasars. More definitive conclusions can
be reached only with larger and more complete samples.

\section{Discussions}

The SMBH masses of AGNs are important to understand the nature of AGN activities, however,
there are only very limited methods which can be used to derive the mass of them
(Ho 1999). 
Many observational studies have shown that a lot of AGNs, especially radio-loud AGNs,
have massive elliptical hosts. The basic properties of these host galaxies are 
indistinguishable with the normal ellipticals.
By assuming 
that the
elliptical hosts of AGNs follow the same fundamental plane as normal ellipticals and adopting
the $M_{BH}$-$\sigma$ relation recently discovered for both inactive and active galaxies, 
we estimated the SMBH masses for 63 BL Lac objects and 29 other AGNs which has been imaged
by HST recently. Our results, though with substantial uncertainties,  
show that the SMBH masses of these AGNs are mostly in the range
of $10^{7.5}M_\odot$ to $10^{9.5}M_\odot$. There are no significant differences in SMBH masses
for different AGNs with elliptical hosts. This seems to be a natural consequence if we believe
that that the tight correlations between the SMBH masses and the galaxy properties
also exist for the host galaxies of AGNs. In Figure 7, we compare our results obtained for 
AGNs with the measured central velocity dispersions and black hole masses of 20 nearby 
elliptical galaxies compiled by Kormendy \& Gebhardt (2001). 
Except for the outlier NGC~4486B whose
 outer region may have been stripped away in the tidal interactions with a
more massive companion galaxy (Faber 1973), the relations of our derived central velocity dispersions and 
SMBH masses with the R-band absolute magnitudes are consistent with the trends for normal
ellipticals. In fact, the derived SMBH masses of AGNs, 
have a significant overlap
with those of massive normal ellipticlas. Moreover, 
we note that the consistency of our derived $\sigma$-$M_R$ relation with the
recent measurement of the stellar velocity dispersion
of Mrk~501 (Barth et al. 2002)  also supports that using the fundamental plane to derive
the central velocity dispersions of AGN elliptical hosts is possible and practical.

%
%                                                One column figure
%----------------------------------------------------------- S_vib
   \begin{figure}
   \centering
   \includegraphics[width=8.7cm, height=10cm]
{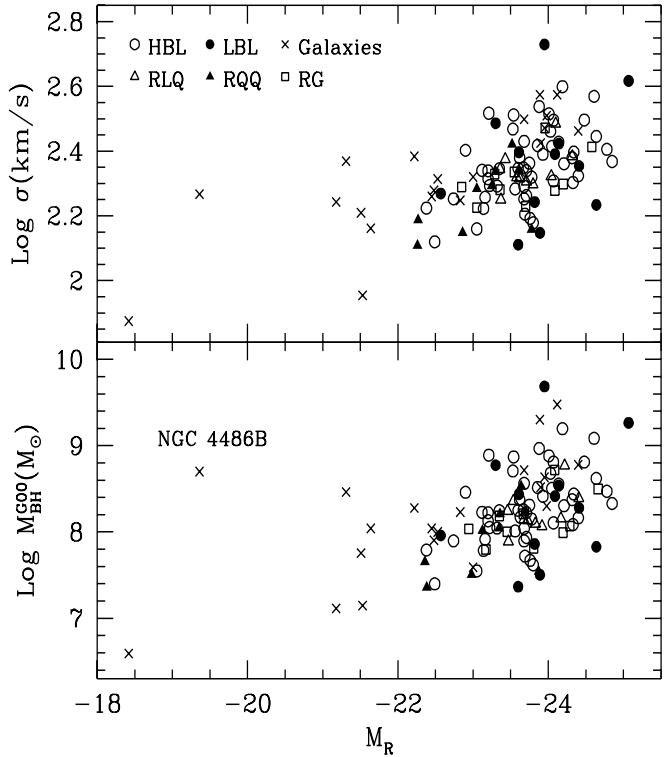}
      \caption{Comparison of the relations of central velocity dispersion, black hole mass 
estimated by the G00 relation with
             the R-band absolute magnitude for AGN hosts and nearby galaxies.}
         \label{Fig7}
   \end{figure}
%
%______________________________________________________________

Our results show that both HBLs and LBLs have the similar SMBH masses in the range
from   $10^{7}M_\odot$ to $10^{9.5}M_\odot$. 
The SMBH masses of BL Lac objects are similar as those in radio galaxies and radio-loud 
quasars. This indicates that the different observational
appearances among HBLs, LBLs  and radio-loud AGNs can not be dominated by the different 
SMBH masses. Several evolutionary scenarios have been recently suggested for radio-loud
AGNs. Ghisellini et al (1998) proposed an evolutionary sequence HBL $\rightarrow$
LBL$\rightarrow$ flat-spectrum radio quasars (FSRQ) according to the increasing level of
external radiation to the soft radiation field in the emitting region. D'Elia \& Cavaliere
(2000) and Cavaliere \& D'Elia (2001) suggested that the gradual depletion of the central
environment may lead to the evolution from FSRQ$\rightarrow$ LBL $\rightarrow$ HBL. 
Similarly, B\"ottcher \& Dermer (2002) recently argued that the decline of accretion 
power 
can also lead to such an evolutionary sequence. As indicated by  B\"ottcher \& Dermer 
(2002), the key parameter of 
these scenarios is  accretion rate rather than the SMBH mass. The depletion of accretion
power can lead to the decrease of the accretion rate, which may result in
the transition of different accretion modes. Theoretical investigations have pointed
out that accretion  near the Eddington limit may
produce optically thick accretion disks extending all the way to the innermost
stable orbit, while accretion at very lower accretion rate may lead to advection
dominated accretion mode (see Narayan et al. 1998 for a review). In fact, 
Fabian \& Rees (1995) has proposed that
the nearby galaxies can be modeled with the advection-dominated accretion flow.
 The 
radio properties of some low-luminosity AGNs may be also related to the advection-dominated
accretion mode (Ulvestad \& Ho 2001). Recently Ghisellini \& Celotti (2001) also proposed
that the separation of FR I and FR II radio galaxies may be closely related to the
critical accretion rates. In addition, the state transition of Galactic black hole
X-ray transients has been explained according to the different accretion modes
at the different accretion rates (Esin et al. 1998).  Recently Fender \& Kuulkers
(2001) also found that the formation and luminosity of jets in
Galactic X-ray transients are closely related to the accretion rates. 
From all these points we suspect that
the main reason for the evolutionary sequence of radio-loud AGNs may be accretion rate rather
than the SMBH mass.
Our estimations of the  SMBH masses and Eddington ratios of different AGNs are also 
consistent with this suspension. 
In addition, our results show that there may be still a difference in SMBH masses of
radio-loud and radio-quiet quasars, but such a difference is not significant as previously claimed.
The Eddington ratios of these two sub-class of quasars seem to be from the same population. These
points are very important to our understanding of the physics of quasars and are obviously need to be 
confirmed with larger samples.

We note that the relations between SMBH mass and the elliptical host  luminosity derived
by us for AGNs are a little different from those obtained by some previous studies. McLure
\& Dunlop  (2002) derived such a relation for 92 active and inactive galaxies
as $\log M_{BH} \propto
-0.50 M_R$.  They argued that it is consistent with a linear scaling between 
the black hole and bulge mass. Wandel (2002) reached a similar conclusion for a sample of 35
quiescent galaxies and 47 broad line AGNs. However, Laor (2001) and Wu \& Han (2001)
obtained steeper slopes of the $M_{BH}$ - bulge luminosity relation for different
samples of AGNs and argued that
the scaling of the black hole and bulge mass is nonlinear. 
Here we provide an additional argument for this nonlinear relation.
In our present study, the slope of the relation between SMBH mass and the elliptical host  luminosity
is estimated from -0.64 to -1.02 for different samples. This is identical to a nonlinear
relation between the black hole and bulge mass
$M_{BH} \propto M_{bulge}^{(1.27\sim1.95)}$ if the mass-to-light ratio of the host galaxy is taken
to be $M/L \propto L^{0.31}$ (Jorgensen et al. 1996).
In fact, we can obtain this conclusion directly from some existed relations.
From the fundamental plane Eq. (1), the $M_{BH}$ -$\sigma$ relation Eq.(3) or (4),
 and the formula for average brightness (Eq.(2)), we can derive a relation between
the SMBH mass and R-band absolute magnitude of the host galaxy as:
$\log M_{BH} \propto (0.96 \sim 1.21) M_R$. This clearly  indicates a nonlinear relation:
$M_{BH}\propto M_{bulge}^{(1.83\sim 2.31)}$. Such a result is in
well agreement with that in Wu \& Han (2001) who obtained 
$M_{BH}\propto M_{bulge}^{(1.74\pm 0.14)}$. 
However, we noted the large uncertainties in our derived SMBH masses and host galaxy luminosities
may have significant effects on the correlation between SMBH and galactic bulge masses. 
Although the non-linear scaling between
them has been also implied in some theoretical models (e.g. Adams
et al. 2001; Wang et al. 2000), 
more detailed theoretical investigations, 
as well as more high quality imaging observations on larger samples of active and inactive
galaxies, are obviously needed to confirm it.

\begin{acknowledgements}
      We thank the referee, Todd Boroson, for a construcitve referee's report and 
Jiansheng~Chen, Jun~Ma, Hong~Wu, Xiang-Ping~Wu, Xu~Zhou for 
stimulating discussions.
This work was supported by the NSFC (No. 10173001) and the Scientific Fundation for 
Returned Overseas Chinese
Scholars, Ministry of Education, China. This research has made use of the NASA/IPAC Extragalactic 
Database (NED) which is operated by the Jet
             Propulsion Laboratory, California Institute of Technology, under contract with 
the National Aeronautics and Space
             Administration. 
. 
\end{acknowledgements}


\begin{thebibliography}{}

 
\bibitem[2001]{adams01} Adams, F.C., Graff, D.G., \& Richstone, D.O. 2001, ApJ, 551, 31
\bibitem[2002]{barth02} Barth, A.J., Ho, L.C., \& Sargent, W.L.W., 2002, \apj, 566, L13
\bibitem[2001]{bettoni01}Bettoni, D., Falomo, R., Fasano, G., et al. 2001, A\&A, 380, 471


\bibitem[2002]{bottcher} B\'ottcher, M., Dermer, C.D., 2002, \apj, 564, 86

\bibitem[2001]{cavaliere} Cavaliere, A., \& D'Elia, V. 2001, \apj, submitted (astro-ph/0106512)
\bibitem[2000]{Delia}D'Elia, A., \& Cavaliere, A. 2000, \pasp, 227, 252
\bibitem[1987]{djorgovski87} Djorgovski, S., Davis, M. 1987, \apj, 313, 59
\bibitem[1987]{dressler87} Dressler, A., Lynden-Bell, D., Burstein, D., et al. 1987, \apj, 313, 42
\bibitem[2002]{dunlop02} Dunlop, J.S., McLure, R.J., KuKula, M.J., \etal 2002, \mnras, 
in press (atro-ph/0108397)
\bibitem[1998]{esin} Esin, A.A., Narayan, R., Cui, W. Grove, J.E., Zhang, S.N. 1998, \apj, 505, 854

\bibitem[1973]{faber73} Faber, S.M. 1973, ApJ, 179, 423
\bibitem[1989]{faber89} Faber, S.M., Wegner, G., Burstein, D., et al. 1989, ApJSS, 69, 763

\bibitem[1995]{fabian} Fabian, A.C., Rees, M.J. 1995, MNRAS, 277, L55




\bibitem[1999]{fan} Fan, J.H., Xie, G.Z., \& Bacon, R. 1999, A\&AS, 136
\bibitem[2001]{fender} Fender, R.P., \& Kuulkers, E. 2001, \mnras, 324, 923

\bibitem[2000]{ferrarese00} Ferrarese, L., \& Merritt, D. 2000, ApJ, 539, L9
\bibitem[2001]{ferrarese01} Ferrarese, L., \etal 2001, ApJ, 555, L79
\bibitem[2000]{gebhardt00a} Gebhardt, K. \etal 2000a, ApJ, 539, L13(G00)
\bibitem[2000]{gebhardt00b} Gebhardt, K. \etal 2000b, ApJ, 543, L5

\bibitem[2000]{ghisellini} Ghisellini, G., Celloti, A. 2001, \aa, 379, L1
\bibitem[1998]{ghisellini} Ghisellini, G., Celloti, A., Fassati, G., Maraschi, L., Comastri, A., 1998, \mnras, 301, 451

\bibitem[1999]{ho99} Ho, L.C. 1999, in Observational Evidence for Black Holes in the Universe,
	ed. S.K. Charkrabarti (Dordrecht: Kluwer), 157
\bibitem[2002]{ho02} Ho, L.C. 2002, \apj, 564, 120
\bibitem[1990]{isobe} Isobe, T., Feigelson, E., Akritas, M.G., Babu, G.J. 1990, \apj, 364, 104
\bibitem[1995]{jorgensen95} Jorgensen, I., Frank, M., Kjaergaard, P. 1995, \mnras, 276, 134
\bibitem[1996]{jorgensen96} Jorgensen, I., Frank, M., Kjaergaard, P. 1996, \mnras, 280, 167

\bibitem[2000]{kaspi00} Kaspi, S., Smith, P.S., Netzer, H., Maoz, D., Jannuzi, B.T., 
	\& Giveon, U. 2000, ApJ, 533, 631
\bibitem[2001]{kormendy01} Kormendy, J. \& Gebhardt, K. 2001, Proceedings of The 20th 
Texas Symposium on 
Relativistic Astrophysics, ed. H. Martel \& J.C. Wheeler, in press (astro-ph/0105230)
\bibitem[2001]{krolik} Krolik, J.H. 2001, \apj, 551, 72
\bibitem[2000]{laor00} Laor, A. 2000, ApJ, 543, L111

\bibitem[2001]{laor01} Laor, A. 2001, ApJ, 553, 677
\bibitem[1998]{magorrian} Magorrian, J., Tremaine, S. Richstone, D., \etal 1998, AJ, 115, 2285
\bibitem[2001]{mclure01} McLure, R.J,  Dunlop, J.S. 2001, MNRAS, 327, 199
 \bibitem[2002]{mclure02} McLure, R.J,  Dunlop, J.S. 2002, MNRAS, in press (astro-ph/0108417)
\bibitem[1999]{mclure99} McLure, R.J, Kukula, M.J., Dunlop, J.S., Baum, S.A., O'Dea, C.P.,
	\& Hughes, D.H. 1999, MNRAS, 308, 377
\bibitem[2001]{merritt01} Merritt, D., \& Ferrarese, L. 2001, \apj, 547, 140 (MF01)

\bibitem[1996]{nelson95} Nelson, C.H., \& Whittle, M. 1995, ApJS, 99, 67
\bibitem[1997]{netzer} Netzer, H., \& Peterson, B.M. 1997, in Astronomical Timing
Series, ed. D. Maoz, A. Sternberg, \& E. Leibowitz (Dordrecgt: Kluwer), 85
\bibitem[1999]{nilsson} Nilsson, K., Pursimo, T., Takalo, L.O., \etal \pasp, 111, 1223
\bibitem[2000]{rieger} Riegger, F.M., \& Mannheim, K. 2000, \aa, 359, 948
\bibitem[1999]{scarpa99} Scarpa, R., Urry, C.M., Falomo, R., \etal 1999, ApJ, 521, 134 
\bibitem[2001]{ulvestad} Ulvestad, J.S., \& Ho, L.C. 2001, \apj, 562, L113

\bibitem[2001]{urry01} Urry, C.M., Scarpa, R., O'Dowd, M. \etal 2000, 532, 816
\bibitem[2002]{wandel02} Wandel, A., 2002, ApJ, 565, 762

\bibitem[1999]{wandel99} Wandel, A., Peterson, B.M., \& Malkan, M.A.  1999, ApJ, 526, 579
\bibitem[2000]{wang} Wang, Y.P., Biermann, P.L., \& Wandel, A. 2000, A\&A, 361, 550
 
\bibitem[2001]{wuhan} Wu, X.-B., Han, J.L., 2001, A\&A, 380, 31
\end{thebibliography}
\end{document}